\documentclass{iopart}
\usepackage{iopams}
\usepackage{subfig}
\usepackage{graphicx}

\begin{document}

\title[\mbox{3-D} imaging in VMI]{Momentum space tomographic imaging of photoelectrons}

\author{ C Smeenk$^1$, L Arissian$^{1, 2}$, A Staudte$^1$, D M Villeneuve$^1$ and P B Corkum$^1$}

\address{$^1$ Joint Laboratory for Attosecond Science, University of Ottawa and National Research Council, 100 Sussex Drive, Ottawa, Canada}
\address{$^2$ Department of Physics, Texas A \& M University, College Station, USA}
\ead{christopher.smeenk "at" nrc.ca}

\begin{abstract}
	We apply tomography, a general method for reconstructing \mbox{3-D} distributions from multiple projections, to reconstruct the momentum distribution of electrons produced via strong field photoionization. The projections are obtained by rotating the electron distribution via the polarization of the ionizing laser beam and recording a momentum spectrum at each angle with a \mbox{2-D} velocity map imaging spectrometer.  For linearly polarized light the tomographic reconstruction agrees with the distribution obtained using an Abel inversion.  Electron tomography, which can be applied to any polarization, will simplify the technology of electron imaging.  The method can be directly generalized to other charged particles.
\end{abstract}

\section{Introduction}
Photoelectron spectroscopy is an important method to study atoms \cite{PhotoIonizationNobleGases}, molecules \cite{ShigemasaN2} and condensed matter \cite{PhotoIonizationSolids}. The photoelectron energy, momentum and angular distributions can identify the state of the residual ion \cite{Czasch2005PRL}, provide insight into correlation effects \cite{Akoury2007Science}, or yield structural information \cite{Wernet2004Science, LIEDMeckel}. Time resolved photoelectron spectroscopy allows us to study the dynamics of these observables in evolving systems \cite{GessnerDissociation}. The technology for measuring electrons is very well advanced. Single electrons can be detected with $\sim$50\% efficiency using electron multiplier tubes and microchannel plates (MCPs). To measure the three dimensional electron momentum distribution a number of approaches have been developed. For example, an array of single channel electron multipliers can obtain  two dimensional information. The time-of-flight then adds the third dimension \cite{L3detector}. A single detector can replace the array if the electron is captured in an electric and magnetic field and recorded with a combination of MCP and position sensitive detector. When used with a multihit capable delayline detector, the complete \mbox{3-D} momentum distribution of single electrons can be measured \cite{COLTRIMSreview}. However, this powerful method is experimentally demanding and comes at the cost of reduced resolution. Furthermore, the \mbox{3-D} momentum of every single electron is not always necessary since the accumulated electron distribution in the laboratory frame is often a sufficient observable.

If an MCP is used in conjunction with a phosphor screen and a CCD camera, the technological side is greatly simplified, and provides, in continuous operation, easy access to the lab frame distributions of electrons and ions. In a widely employed detection scheme known as velocity map imaging (VMI) \cite{EppinkAndParker}, only electrostatic fields are used to guide the electrons to the detector. Through an integrated electrostatic lens a direct mapping of the initial photoelectron velocity distribution onto the detector is achieved. To obtain a 4$\pi$ collection angle these electric fields are typically several kV in strength resulting in time-of-flight distributions on the order of the temporal response of the detector. Therefore, in VMI the information along the axis perpendicular to the detector is lost.

Previous approaches to determine the \mbox{3-D} momentum distribution in VMI experiments used inverse Abel transforms \cite{FourierHankelInversion, VrakkingAbel, BASEX}, slice imaging \cite{Townsend2003}, or time-resolved event counting \cite{Strasser2000, Dinu2002}. Each of these techniques suffers limitations when applied to non-symmetric distributions. By construction the Abel methods are limited to cases with cylindrical symmetry. The other methods can be used to image non-symmetric distributions in \mbox{3-D} provided the particles are suitably spread in time-of-flight. For electrons this is difficult and new approaches are needed.

A recent attempt to address this problem was to use a tomographic method to reconstruct the momentum distribution in three dimensions \cite{KasselTomography, SmeenkMSc}. This allows the reconstruction of a non-symmetric distribution in a high data throughput experiment like VMI.

We build on the idea of Wollenhaupt \emph{et al} \cite{KasselTomography} and show how this technique can be applied to tunnel ionization of an atom. In experiments with strong laser fields, control of the polarization is essential for affecting electron re-collision processes on attosecond time scales. Imaging the \mbox{3-D} photoelectron distribution in arbitrarily polarized light therefore provides useful information for strong field science. We apply the filtered backprojection technique widely used in medical CT to reconstruct the electron distribution. This is in contrast to Wollenhaupt \emph{et al} who used a Fourier approach. The filtered backprojection technique can give exact reconstructions compared to approximate results obtained with unfiltered techniques \cite{SheppLogan}. Using multiphoton ionization of argon as an example, we make a quantitative comparison between the tomographic method and the Abel inversion method for the case where an Abel inversion is possible. We present the reconstructed \mbox{3-D} electron momentum distributions created in linearly and elliptically ($E_1/E_2= 0.89$)  polarized light in the tunnelling regime.

\section{Methods}
\label{sec:tomographicMethod}
\begin{figure}
	\centering
	\subfloat[Principle of velocity map imaging.]{
		\includegraphics[width=0.45\linewidth]{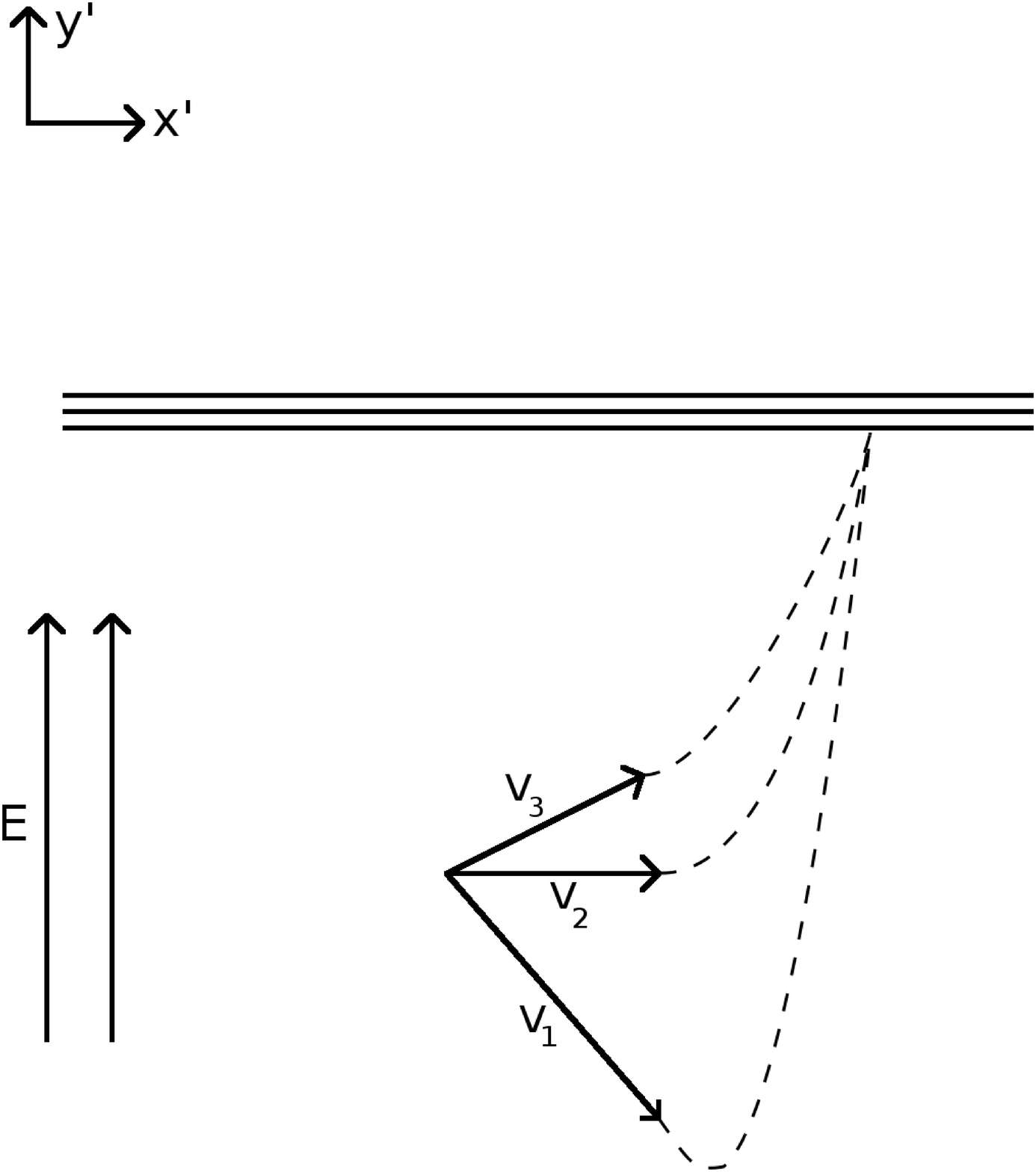}
		\label{fig:principleVMI}
	}
	\subfloat[Principle of tomographic imaging.]{
		\includegraphics[width=0.45\linewidth]{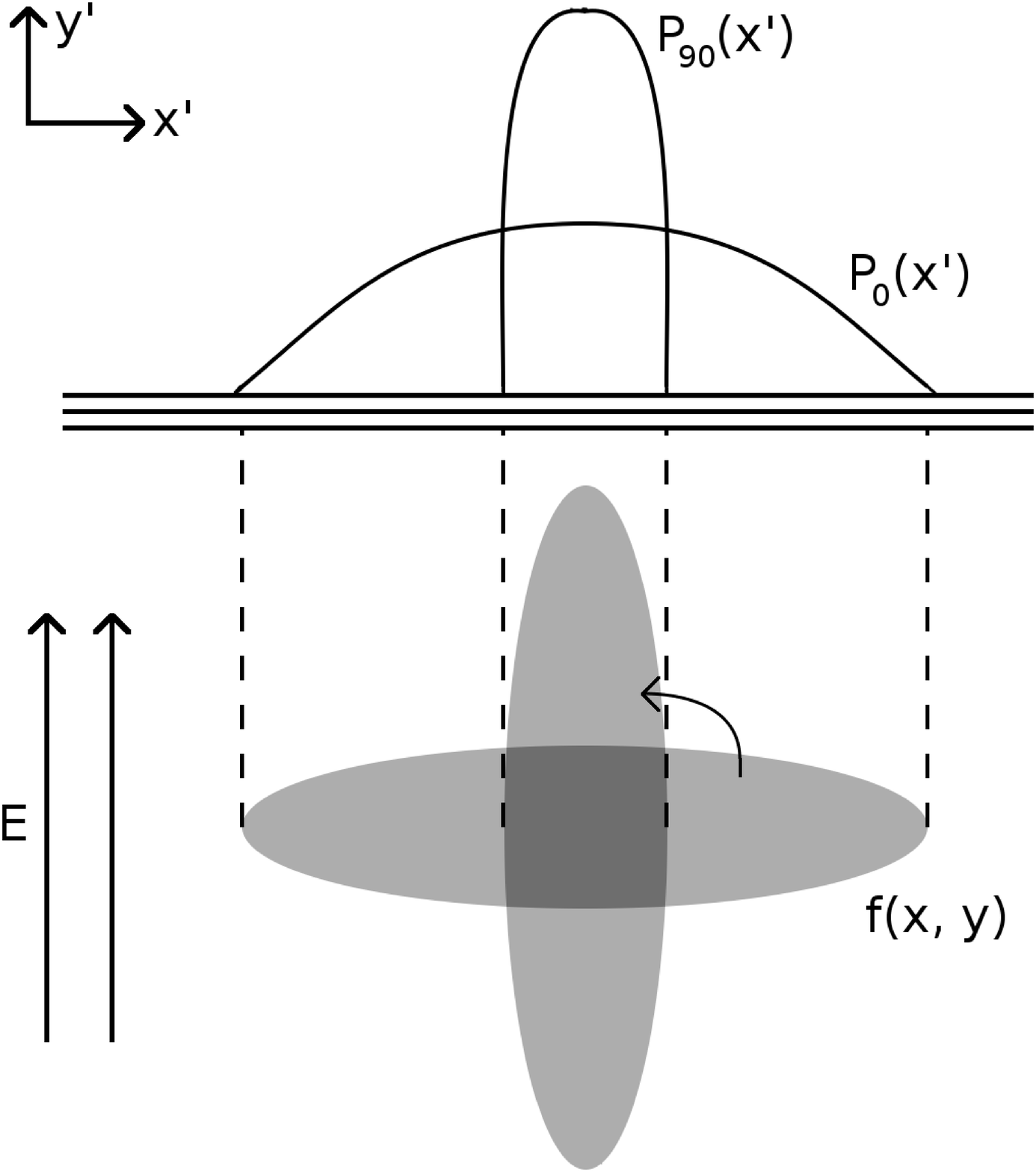}
		\label{fig:principleTomo}
	} \\
	\subfloat[Experimental setup for VMI tomography.]{
		\includegraphics[width=0.5\linewidth]{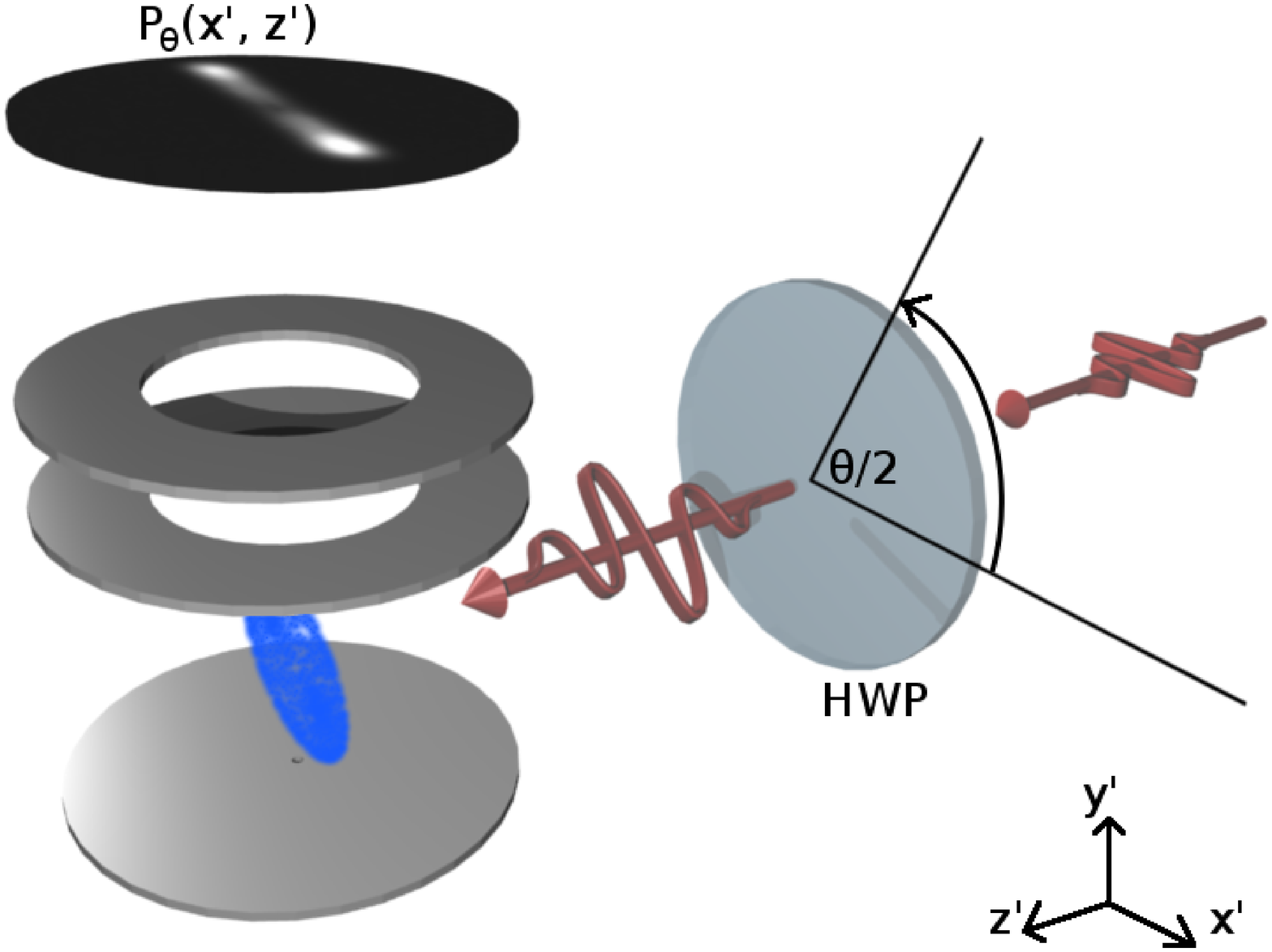}
		\label{fig:tomoSetup}
	}
	\caption{ (a) The principle of VMI. An electrostatic lens projects the $x'$ component of the velocity vector (dark arrows) into the plane of the detector. (b) The principle of tomographic imaging. A distribution $f \left( x , y \right)$ is shown at two different orientations and projected onto the detector along the electric field $E$. (c) Experimental setup for VMI tomography. See text for details. }
\end{figure}

\subsection{Velocity map imaging}
\label{sec:principleVMI}
In velocity map imaging experiments \cite{EppinkAndParker}, an inhomogeneous electric field is used to project the velocity vector of charged particles into the \mbox{2-D} plane of the detector. This is shown schematically in figure \ref{fig:principleVMI}. The electric field acts as a lens which `pancakes' the \mbox{3-D} velocity distribution, preserving it's magnitude along the $x'$ axis and integrating over the $y'$ axis. The vectors $v_1, v_2, v_3$ all having the same $v_{x'}$ but different $v_{y'}$ are projected onto the same spot on the detector. For this to work it is necessary that the initial velocity of the particles is negligible compared to their acceleration in the spectrometer's DC field. The $y'$ component is not observable unless additional information is provided, such as symmetry assumptions or, as we will show, measurements from multiple directions. The former is a condition for Abel inversion and allows the inversion of a single projection. The latter is a requirement for tomographic inversion and is a more general approach.

\subsection{Tomographic imaging}
The principle behind tomographic imaging \cite{KakSlaney} is sketched in figure \ref{fig:principleTomo}. An arbitrary distribution $f \left(x, y \right)$ is shown at two different orientations relative to a detector (shown as the horizontal lines). 
The detector frame of reference $\left( x', y', z' \right)$ is related to the distribution frame of reference by,
\begin{equation}
	\left( \begin{array}{c} x' \\ y' \\ z' \end{array} \right) = \left( \begin{array}{ccc} \cos \theta & - \sin \theta & 0 \\ \sin \theta & \cos \theta & 0 \\ 0 & 0 & 1 \end{array} \right) \left( \begin{array}{c} x \\ y \\ z \end{array} \right) .
	\label{eq:detectorFrame}
\end{equation}
For the application of the tomographic method the rotated distribution must be projected onto the detector. Mathematically this is given by the  Radon transform, 
\begin{equation}
	P_\theta \left(x', z' \right) = P_\theta \left( x', z \right) = \int f \left( x, y, z \right) \; \mathrm{d}y' .
	\label{eq:projection}
\end{equation}
Physically, the Radon transform is carried out in a VMI experiment by the spectrometer's DC electric field, oriented along the long arrows in figure \ref{fig:principleTomo}. The projection of the velocity vector discussed in section \ref{sec:principleVMI} is the experimental Radon transform. Examples of the projections of $f \left(x, y\right)$ at two different orientations are sketched as the curves $P_0 \left( x' \right)$ and $P_{90} \left( x'\right)$ in figure \ref{fig:principleTomo}. By collecting projections for many different orientations the tomographic algorithm is able to reconstruct the complete velocity distribution. Unlike medical tomography, wherein a detector typically covers a portion of the volume of interest, the projections in VMI experiments span the complete momentum space of the photoelectrons. Thus, by rotating the distribution there is enough information to recover the full \mbox{3-D} distribution for any polarization state.

\subsection{VMI tomography}
The setup for VMI tomography is depicted in figure \ref{fig:tomoSetup}. The laser propagates along the $z'$ ($z$) axis with polarization in the $x'y'$ ($xy$) plane. In figure \ref{fig:tomoSetup} the polarization is shown as linear, but in general it can be elliptical. A half wave plate is used to rotate the axes of the polarization. The ability to rotate the polarization and thereby the velocity distribution in the chamber is a necessary step in tomographic reconstruction. In section \ref{sec:results}, results are presented for linear and elliptical polarization.

The beam then enters the VMI spectrometer where it is focused to the centre of the spectrometer using a parabolic mirror (not shown in figure \ref{fig:tomoSetup}). The laser focus is orthogonally crossed by a supersonic gas jet having a density corresponding to $5 \times 10^{-7}$ torr. The VMI electrodes are negatively biased with the correct voltage ratio to project the electron velocity distribution (shown as the dark cloud in figure \ref{fig:tomoSetup}) onto the \mbox{2-D} detector. The electrons are accelerated to roughly $2.5$ keV in the spectrometer's DC field or about 140 times their kinetic energy acquired from the laser pulse. The detector is pulsed by a high voltage  switch to eliminate noise and dark signal. The timing gate on the high voltage pulse is 300 ns long. The amplified signal from the detector is read by a CCD camera acquiring continuously at 12 fps. The Ti:Sa laser system used for these experiments produces 50 fs pulses at 500 Hz centred at 805 nm with a bandwidth of 22 nm.

\subsection{Image inversion}
The inversion of the \mbox{2-D} projections follows the parallel ray filtered backprojection algorithm \cite{KakSlaney}. This method can yield exact reconstructions in contrast to techniques without a filter step. In any computer implementation of the algorithm, the projections are always sampled at some distance interval $\tau$ and some angular interval $\Delta \theta$. The first step in the tomographic algorithm is to filter each measured projection $P_\theta \left(x', z\right)$. This is implemented as the convolution of $P_\theta \left( x' \right)$ with a filter function. We have used the discrete ``Shepp - Logan'' filter \cite{SheppLogan}:
\begin{equation}
	h \left( x' \right) = h \left( n \tau \right) = - \frac{2}{\pi^2} \left( \frac{1}{4 n^2 - 1} \right)
\end{equation}
where $x' = n \tau$ with $\tau$ as the sampling interval and $n$ is an integer. The filtered projection is
\begin{equation}
	Q_\theta \left( x', z \right) = h ( x' ) \ast P_\theta ( x', z ) 
	\label{eq:Qprojection}
\end{equation}
As explained in section \ref{sec:principleVMI}, the data were originally acquired by integration along the spectrometer's electric field ($y'$ axis). At this stage in the reconstruction, each filtered projection $Q_\theta \left( x', z\right)$ must be backprojected along the $y'$ axis. Summing over all the projection angles yields the reconstructed distribution. Mathematically, the reconstructed distribution $F \left( x, y, z \right)$ is given by
\begin{eqnarray}
	 F \left( x, y, z \right) &=& {\displaystyle \int_{0}^{\pi} Q_\theta \left(x', z \right) \; \mathrm{d}\theta } \label{eq:backpCont} \\ 
	 &=& {\displaystyle \frac{\pi}{K}  \sum_{i = 1}^K Q_{\theta_i} \left( x \cos \theta_i - y \sin \theta_i, z \right) } \label{eq:backpDiscreet}
\end{eqnarray}
where the projections are sampled at $K$ different angles and $0 \leq \theta_i \leq \pi$. Implementing the discrete backprojection (\ref{eq:backpDiscreet}) requires one-dimensional interpolation to find the projection value $Q_{\theta_i} \left(x' \right)$ at each point $\left( x, y \right)$ in the distribution frame of reference.

Equation (\ref{eq:backpDiscreet}) is the reconstructed momentum distribution in three dimensions. Thus, by rotating the laser polarization the tomographic method allows us to reconstruct the complete momentum distribution.

\section{Results}
\label{sec:results}
\begin{figure}
	\centering
	\includegraphics[width=0.7\linewidth]{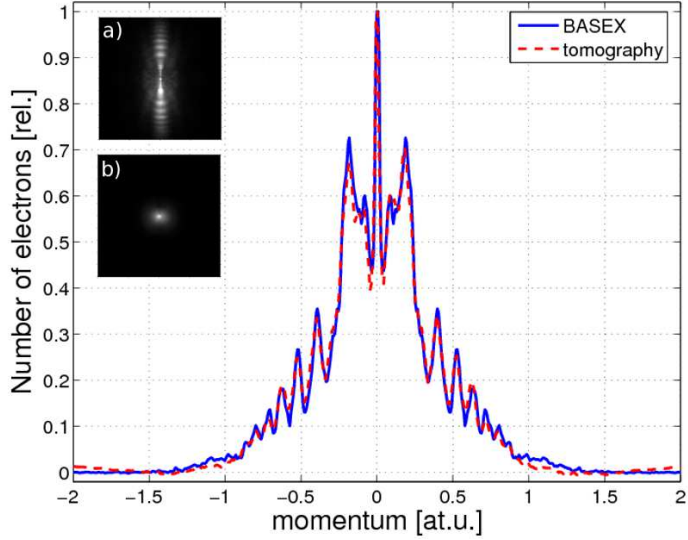}
	\caption{Profile along the polarization axis of the \mbox{3-D} momentum distribution. Blue (solid) line: distribution inverted using the inverse Abel (BASEX) method \cite{BASEX}. Red (dashed) line: distribution inverted using the tomographic method. Inset: examples of projections used in the tomographic reconstruction a) $\theta = 0^\circ$, b) $\theta = 90^\circ$}
	\label{fig:AbelTomoLinear}
\end{figure}

First, we compare the tomographic method with the inverse Abel method for linearly polarized light. For an Abel inversion the polarization axis must be parallel to the detector face. A single two-dimensional projection was recorded at a laser intensity of $2 \times 10^{14} \; \mathrm{W/cm}^2$ and then inverted using the inverse Abel method. A profile along the polarization axis in the inverted distribution is shown as the solid line in figure \ref{fig:AbelTomoLinear}. The profile has been normalized to its maximum value. The distribution shows multiple peaks spaced by one  photon's energy. This is characteristic of above threshold ionization \cite{AgostiniATI}.

Next, the tomographic method was tested under the same conditions. The laser parameters were held constant and the polarization direction was rotated using a broadband half-waveplate (Bernhard Halle GmbH). Examples of two projections taken at $0$ and $90$ degrees are shown in figure \ref{fig:AbelTomoLinear}a and \ref{fig:AbelTomoLinear}b. In figure \ref{fig:AbelTomoLinear}a the polarization is parallel to the detector face, and in figure \ref{fig:AbelTomoLinear}b it is perpendicular. The two dimensional projections $P_\theta \left(x', z\right)$ were recorded as the polarization rotated from $0 - 90$ degrees in steps of $2$ degrees. These projections were then duplicated to yield the $\pi$ projection angles required by the parallel ray backprojection algorithm.


The tomographic method was used to reconstruct the momentum distribution in 15 different slices through the \mbox{3-D} momentum distribution. A spacing of $0.071$ atomic units was used between each adjacent slice along the laser propagation ($z$) axis. The result is the \mbox{3-D} electron momentum distribution. A profile along the polarization axis and in the plane of polarization ($z = 0$) is shown as the dashed line in figure \ref{fig:AbelTomoLinear}. Again, the profile has been normalized to its maximum value. The profile generated by the tomographic method is reconstructed using a series of \mbox{1-D} projections measured at different angles. This is in contrast to the inverse Abel method which uses a \mbox{2-D} projection to infer the structure of the momentum distribution. There is good agreement between the tomographic method and the result from the BASEX method \cite{BASEX}. This shows the tomographic method reproduces the result of the inverse Abel transform for linearly polarized light.

\begin{figure}
	\subfloat[ ]{
		\includegraphics[width=0.5\linewidth]{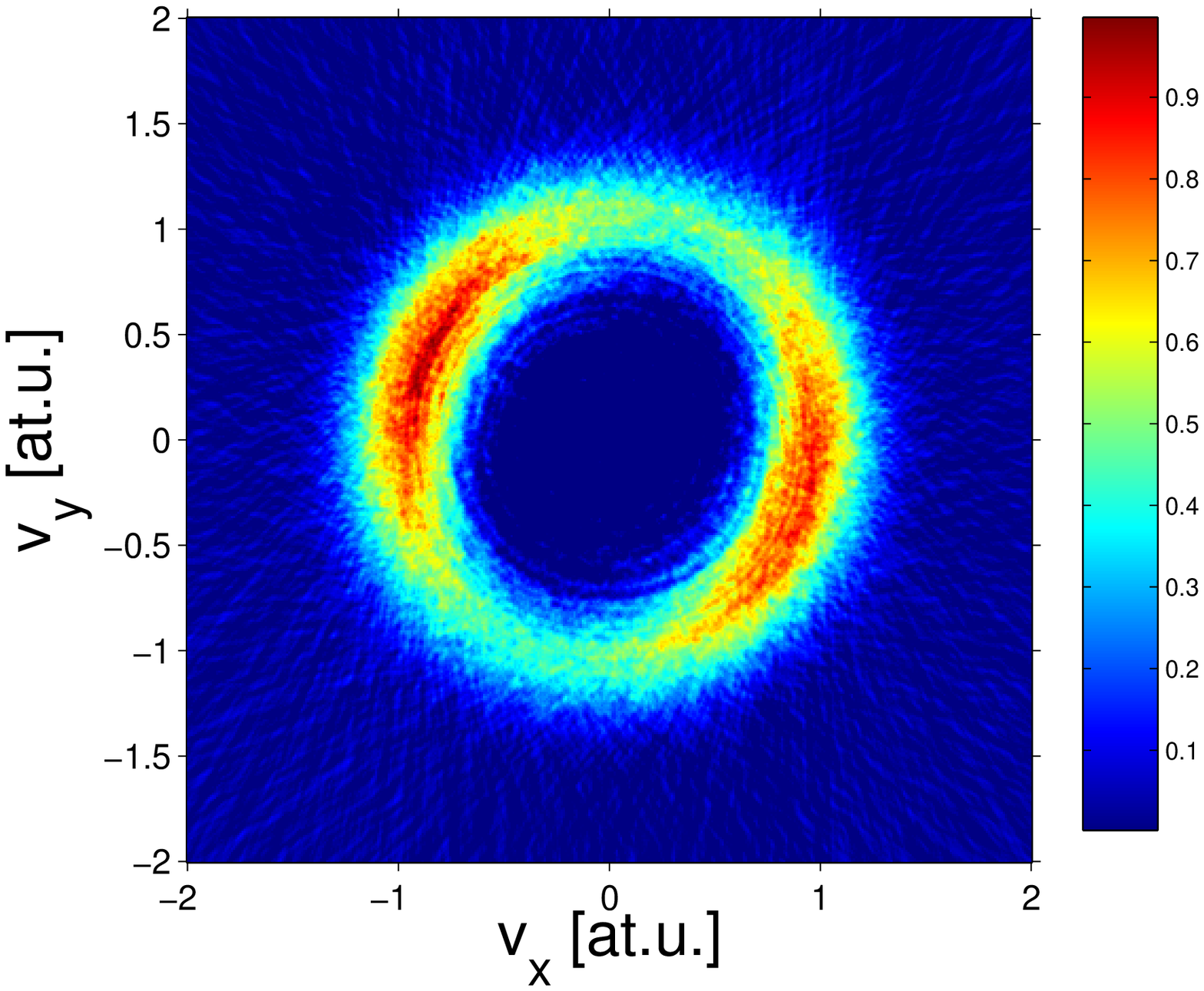}
		\label{fig:donut2D}
	}
	\subfloat[ ]{
		\includegraphics[width=0.5\linewidth]{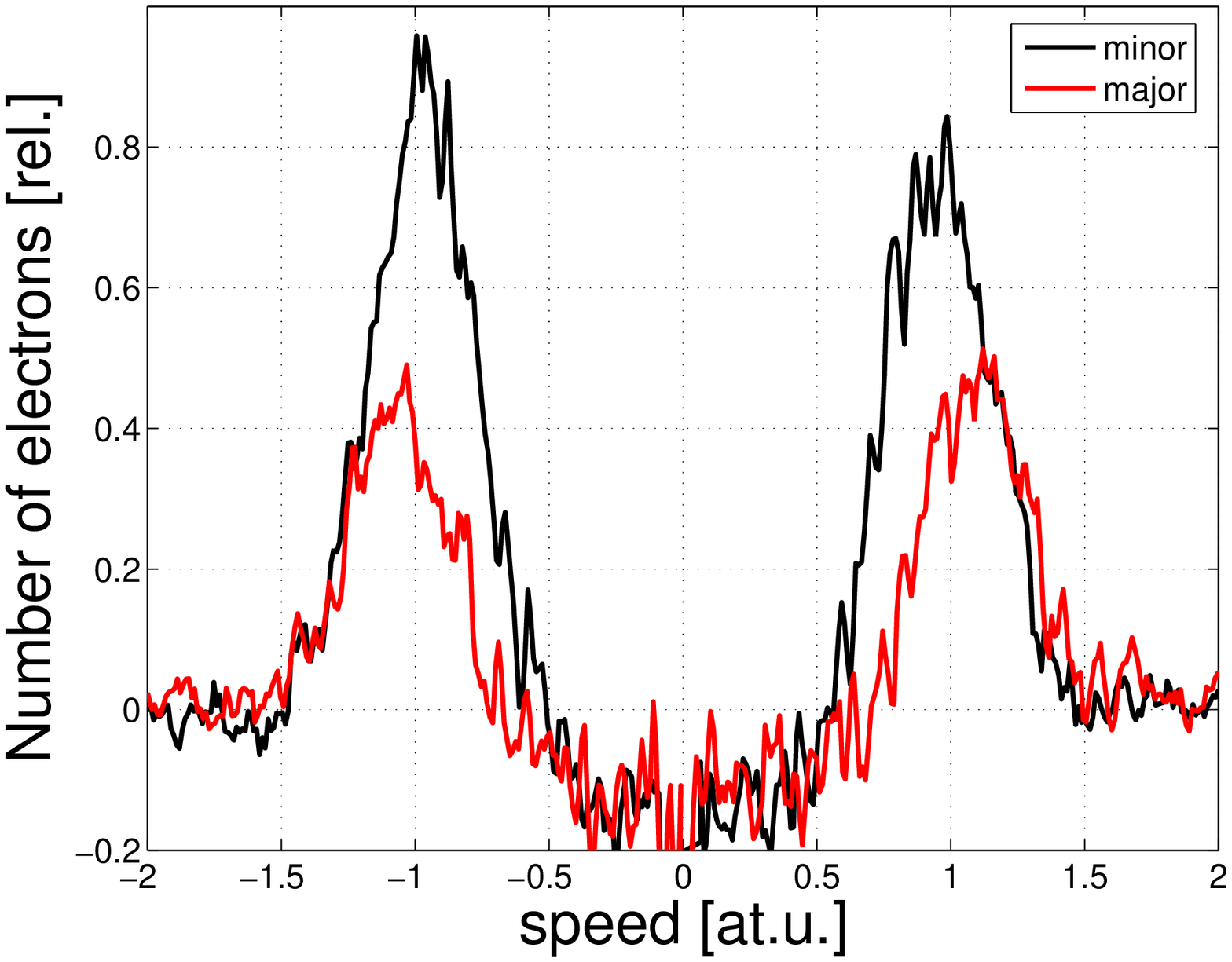}
		\label{fig:donutProfile}
	}
	\caption{Tomographic imaging of the electron velocity spectrum in elliptically polarized light. (a) Reconstructed distribution in the plane of polarization. The colour code represents the relative number of electrons. (b) The number of electrons as a function of their speed measured along the minor (black) and major (red) axes.}
\end{figure}

The power of the tomographic method lies in its application to distributions that need not contain any symmetry. A quarter wave plate was inserted into the beam and rotated by $43^\circ$ to create elliptically polarized light at an intensity of $3 \times 10^{14}$ W/cm$^2$. Using the half wave plate in figure \ref{fig:tomoSetup} the polarization ellipse was rotated from $0 - 178$ degrees in steps of 2 degrees. At each angle a projection of the electron velocity distribution containing roughly $6 \times 10^5$ electrons was acquired. The tomographic method was used as before to reconstruct the \mbox{3-D} velocity distribution. The distribution in the plane of polarization is shown in figure \ref{fig:donut2D}.

Ellipticity appears  in two ways in the reconstructed image. The most obvious evidence is in the nonuniform momentum distribution. This is shown more clearly in figure \ref{fig:donutProfile} where a profile along the major axis is shown as the red line; the minor axis is shown as the black line. In figure \ref{fig:donutProfile} it is clear that there are more electrons along the minor axis of the ellipse. This counterintuitive result is explained in reference \cite{Dietrich2000}. It occurs because the ionization rate is a function of the laser's instantaneous electric field while the drift velocity is a function of the vector potential at the phase of ionization. Since the two are $\pi/2$ out of phase, the larger ionization rate appears along the minor axis of the velocity distribution, as shown in figure \ref{fig:donutProfile}. Using the size of each of the axes, the laser ellipticity is found to be $0.89$.

\begin{figure}
	\subfloat[linear polarization]{
		\includegraphics[width=0.5\linewidth]{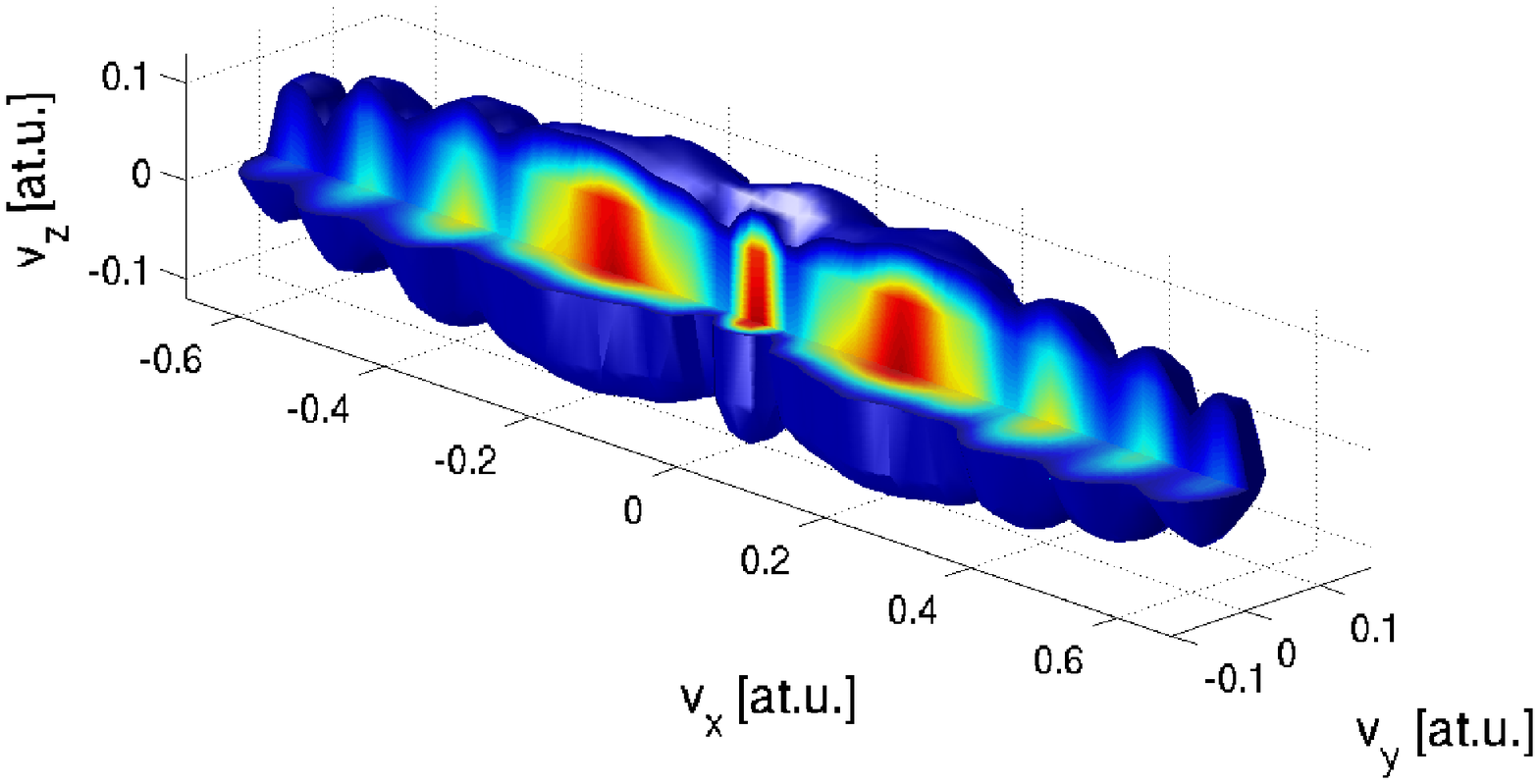}
	}
	\subfloat[elliptical polarization]{
		\includegraphics[width=0.5\linewidth]{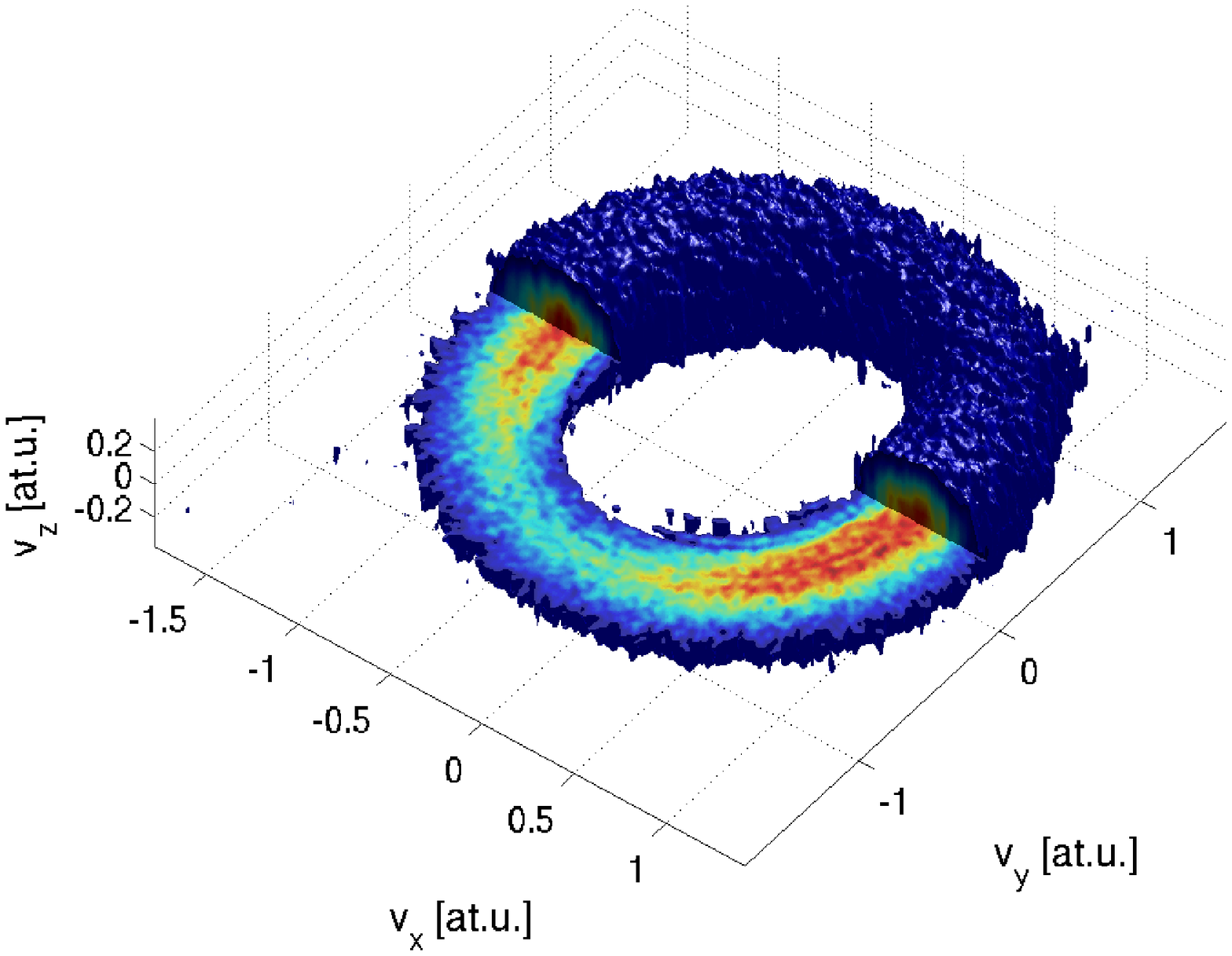}
	}
	\caption{\mbox{3-D} reconstructions using the tomographic method. The colour code is the same as figure \ref{fig:donut2D}. Values less than $0.1$ are transparent.}
	\label{fig:donut3D}
\end{figure}

The tomographic method can be applied to visualize the \mbox{3-D} velocity distribution for any polarization state. Figure  \ref{fig:donut3D} shows a three-dimensional representation of the reconstructed velocity distribution in both linearly and elliptically polarized light. A surface of constant probability is shown in navy blue. On the interior of the distributions the colour coding reflects the probability to measure an electron at each point in the \mbox{3-D} distribution. In figure \ref{fig:donut3D} the data was averaged over neighbouring volume elements.

\section{Conclusion}
We have shown how ideas of tomographic imaging can be extended to study tunnel ionization in velocity map imaging experiments. No assumptions are needed about the symmetry of the measured distribution. The method accurately reproduces the \mbox{3-D} momentum distribution in linearly polarized light. We have applied the method to reconstruct the velocity distribution of electrons in multiphoton ionization of argon by elliptically polarized light -- a case not handled by inversion techniques based on the Abel transform.
%

There have been some recent papers \cite{LimitsToTunnelling, NovelTunnelling} questioning the applicability of tunnelling models \cite{KeldyshTunnel} to strong field phenomena in 800 nm or even 1400 nm laser light. Experiments with circular or elliptical polarization will allow the tunnelling model to be tested without the complexity of re-collision. Resolving the ionized electrons in three dimensions will give new insight into the ionization mechanism.

\ack
We are pleased to acknowledge financial support from NSERC, the Canadian Institute for Photonics Innovations, US Air Force Office of Scientific Research, and MURI grant W911NF-07-1-0475. We are grateful for the help of Domagoj Pavi\v{c}i\'c, Mike Wong and Andrei Naumov in the design and construction of the VMI chamber.

\section*{References}
\bibliographystyle{unsrt} 

\begin{thebibliography}{10}

\bibitem{PhotoIonizationNobleGases}
D~J Kennedy and S~T Manson.
\newblock {\em Phys. Rev. A}, 5(1):227, 1972.

\bibitem{ShigemasaN2}
E~Shigemasa, J~Adachi, M~Oura, and A~Yagishita.
\newblock {\em Phys. Rev. Lett.}, 74(3):359, 1995.

\bibitem{PhotoIonizationSolids}
F~J Grunthaner, P~J Grunthaner, R~P Vasquez, B~F Lewis, and J~Maserjian.
\newblock {\em Phys. Rev. Lett.}, 43(22):1683, 1979.

\bibitem{Czasch2005PRL}
A~Czasch~\emph{et al}.
\newblock {\em Phys. Rev. Lett.}, 95:243003, 2005.

\bibitem{Akoury2007Science}
D~Akoury~\emph{et al}.
\newblock {\em Science}, 318:949--952, 2007.

\bibitem{Wernet2004Science}
Ph~Wernet~\emph{et al}.
\newblock {\em Science}, 304:995--999, 2004.

\bibitem{LIEDMeckel}
M~Meckel~\emph{et al.}
\newblock {\em Science}, 320(5882):1478 -- 1482, 2008.

\bibitem{GessnerDissociation}
O~Ge\ss{}ner~\emph{et al.}
\newblock {\em Science}, 311:219--223, 2006.

\bibitem{L3detector}
A~Adam~\emph{et al}.
\newblock {\em Nucl. Inst. Meth. A}, 383(2-3):342--366, 1996.

\bibitem{COLTRIMSreview}
J~Ullrich~\emph{et al.}
\newblock {\em Rep. Prog. Phys.}, 66(9):1463 -- 1545, 2003.

\bibitem{EppinkAndParker}
A~Eppink and D~Parker.
\newblock {\em Rev. Sci. Instr.}, 68(9):3477 -- 3484, 1997.

\bibitem{FourierHankelInversion}
L~Smith, D~Keefer, and S~Sudharsanan.
\newblock {\em J. Quant. Spectrosc. Rad. Trans.}, 39(5):367 -- 373, 1988.

\bibitem{VrakkingAbel}
M~Vrakking.
\newblock {\em Rev. Sci. Instr.}, 72(11):4084 -- 4089, 2001.

\bibitem{BASEX}
V~Dribinski, A~Ossadtchi, V~Mandelshtam, and H~Reisler.
\newblock {\em Rev. Sci. Instr.}, 73(7):2634 -- 2642, 2002.

\bibitem{Townsend2003}
D~Townsend, M~P Minitti, and A~G Suits.
\newblock {\em Rev. Sci. Instr}, 74(4):2530--2539, 2003.

\bibitem{Strasser2000}
D~Strasser~\emph{et al}.
\newblock {\em Rev. Sci. Instr.}, 71(8):3092--3098, 2000.

\bibitem{Dinu2002}
L~Dinu, Eppink A T~J B, F~Rosca-Pruna, H~L Offerhaus, W~J van~der Zande, and
  Vrakking M~J J.
\newblock {\em Rev. Sci. Instr.}, 2002(12):4206--4212, 2002.

\bibitem{KasselTomography}
M~Wollenhaupt, M~Krug, J~Kohler, T~Bayer, C~Sarpe-Tudoran, and T~Baumert.
\newblock {\em App. Phys. B}, 95:647--651, 2009.

\bibitem{SmeenkMSc}
C~Smeenk.
\newblock Velocity map imaging of electrons generated via femtosecond laser
  ionization.
\newblock Master's thesis, University of Ottawa (Canada), Feb.\ 2009.

\bibitem{SheppLogan}
L~Shepp and B~Logan.
\newblock {\em IEEE Trans. Nucl. Sci.}, NS-21(1):21 -- 43, 1974.

\bibitem{KakSlaney}
A~Kak and M~Slaney.
\newblock {\em Principles of Computerized Tomographic Imaging}.
\newblock IEEE Press, 1988.

\bibitem{AgostiniATI}
P~Agostini, F~Fabre, G~Mainfray, G~Petite, and N~K Rahman.
\newblock {\em Phys. Rev. Lett.}, 42(17):1127--1130, 1979.

\bibitem{Dietrich2000}
P~Dietrich, F~Krausz, and P~B Corkum.
\newblock {\em Optics Lett.}, 25(1):16--18, 2000.

\bibitem{LimitsToTunnelling}
H~R Reiss.
\newblock {\em Phys. Rev. Lett.}, 101(4):043002, 2008.

\bibitem{NovelTunnelling}
H~R Reiss.
\newblock {\em Phys. Rev. Lett.}, 102(14):143003, 2009.

\bibitem{KeldyshTunnel}
L~V Keldysh.
\newblock {\em Sov. Phys. JETP}, 20:1307--1314, 1965.

\end{thebibliography}

\end{document}